\documentstyle[12pt]{article}
\newcommand{\beq}{ \begin{equation} }
\newcommand{\eeq}{ \end{equation} }
\begin{document}
\baselineskip24pt

\begin{center}
{\Large {\bf Coarse graining: lessons from simple examples}} 
\vskip0.5cm
{\bf P.~Akritas$^{1,2}$, I.~Antoniou$^{1,2}$, E. Yarevsky$^{1,3}$}
\end{center}

\vskip0.3cm \noindent 
$^1${\it International Solvay Institutes for Physics
and Chemistry, Campus Plaine ULB C.P.231, Bd.du Triomphe, Brussels 1050,
Belgium} \newline
$^2${\it Theoretische Natuurkunde, Free University of Brussels, C.P.231,
Brussels 1050, Belgium} \newline
$^3${\it Laboratory of Complex Systems Theory, Institute for Physics,
St.Petersburg State University, Uljanovskaya 1, Petrodvoretz, St.Petersburg
198904, Russia} \noindent

\vskip1cm
\begin {abstract}
We assess Coarse Graining by studying different partitions of the phase
space of the Baker transformation and the periodic torus automorphisms. 
It turns out that the shape of autocorrelation functions for the Baker
transformation is more or less reproduced. However, for certain partitions 
the decay rates turn out to be irrelevant, even decay may stop in a finite
time. For the periodic torus automorphisms, Coarse Graining introduces
artificial dumping.
\end {abstract}



\vskip5mm
Corresponding author:\\
Prof. I. Antoniou\\
International Solvay Institutes for Physics and Chemistry, 
Campus Plaine ULB C.P.231, Bd.du Triomphe, Brussels 1050, Belgium \\
Phone: +32-2-6505048 \\
Fax: +32-2-6505028\\
E-mail: iantonio@vub.ac.be

\newpage
\section {Introduction}

Coarse graining is a simple way to explain the manifest irreversibility 
from the underlying entropy preserving dynamical laws. The idea which
goes back to Ehrenfests~\cite {Ehrenfest}, see also Tolman~\cite {Tolman},
is that fine measurements at the fundamental level are unattainable,
therefore we can and should observe only averages over microstates.
The averaging is introduced by the observer in addition to the dynamical 
evolution. The resulting loss of information gives rise to entropy 
increase. In fact all other conventional explanations of irreversibility 
are inventions of extra-dynamical mechanisms to lose dynamical information.
Boltzmann's Stossalanzats~\cite {Ehrenfest,Boltzmann} amounts to loss of
postcollisional correlations, von Neumann's~\cite {vonNeumann} measurement
projection postulate amounts to loss of quantum correlations described by 
the off-diagonal matrix elements of the representation of the pure state in 
terms of a basis of common eigenvectors of a complete system of commuting
observables, resulting in the collapse of the wave function. Decoherence
amounts to loss of information through transfer to an unknown and
uncontrollable environment~\cite {Zurek}.

We shall not discuss here the details of these approaches which involve 
assumptions additional to the dynamical evolution of a more or less 
subjectivistic character. For a recent discussion see for example~\cite 
{Lebowitz}. In contradistinction Prigogine and coworkers have 
stressed~\cite {Prig7,Prig8,Prig9} that irreversibility should be an 
intrinsic property of dynamical systems which admit time asymmetric 
representations of evolution, without any loss of information. These 
time asymmetric representations can be obtained by extending the evolution
to suitable Rigged Hilbert Space or by intertwining the evolution with
Markov processes. These intrinsically irreversible systems include for
example Large Poincar\'e Non-Integrable systems and chaotic systems.
We shall not discuss further these interesting direction of research. 
Our objective is to see how coarse graining works for simple systems 
where calculations are controllable and then draw more general conclusions.
In Section~2 we define the coarse graining projections and the coarse 
grained evolution. We study coarse graining projection of the Baker
transformation in Section~3 and of the periodic torus automorphisms 
in Section~4.

\section {Coarse graining projections}

We introduce the basic concepts and notations that will later be used in
the description of particular systems. Let us consider the configuration
space $X$ with the measure $\mu $. The partition $\zeta$ is the finite set 
$\{Z_{k}\}_{k=1}^{M}$ of the cells $Z_{k}$ which satisfy the
following properties:
\beq
\begin{array}{l}
\bigcup_{k=1}^{M}Z_{k}=X, \\
\mu (Z_{i}\bigcap Z_{k})=\delta_{ik}\mu (Z_{i}),
\end{array}
\eeq
where $\delta _{ik}$ is the Kroneker symbol.

Coarse graining is implemented through the averaging projection over 
the cells of the partition $\zeta$ known also in probability 
theory~\cite {ref10,ref11} as the conditional expectation operator $P$
over $\zeta$:
\begin{equation}
\label {proj-oper}
Pf(x)=\sum_{k=1}^{M}f_{k}\ 1_{Z_{k}}(x).
\end{equation}
Here $1_{Z_{k}}(x)$ is the indicator of the set $Z_{k}$:
$$
1_{Z_{k}}(x)= \left \{
\begin{array}{ll}
1, & x\in Z_{k} \\
0, & x\notin Z_{k}
\end{array}
\right. ,
$$
and the average value $f_{k}$ of the function $f(x)$ in the cell $Z_{k}$ is
$$
f_{k}={\frac{1}{\mu (Z_{k})}}\int_{Z_{k}}d\mu (x)f(x).
$$
Let us define by $Q$ the orthocomplement of $P$:
$$
P+Q=I.
$$
$Q$ projects onto the fine/detail information eliminated by $P$.
The simplest dynamics are cascades of automorphisms $S$ of the phase space
$X$. The observables are usually square integrable phase functions in
$L^2(X)$ and they evolve according to the iterated Koopman operator
$$
V: L^2(X) \rightarrow L^2(X) 
\quad \mbox{such that} \quad 
Vf(x) = f(Sx).
$$
The coarse grained evolution is described by the projected evolution
by $P$~(\ref{proj-oper}):
\beq
V_P^n = (PVP)^n, 
\quad n =0,1,2\ldots .
\eeq

In general however
\beq
VP = PVP+QVP.
\eeq
Therefore starting with a coarse grained observable $Pf$, the evolution
may regenerate the fine detail $QVPf$, which is eliminated by the repeated
application of the projection $P$. Therefore arbitrary projections $P$ 
destroy the dynamical evolution. In fact as the Koopman operators of
chaotic systems are shifts~\cite {Foias}, their coarse grained projections
can be whatever contractive evolution one wishes (Structure theorem of
bounded operators)~\cite {Foias,Filmore} with any desired decay rates.
A condition which guarantees the reliability of the
coarse grained description is that the evolution $V$ should not destroy 
the cells of the coarse graining partition. In this way minimal dynamical
information is lost by averaging and the resulting symbolic dynamics is
very close to the original evolution. Such coarse-graining
projections compatible with the dynamics should be distinguished from
arbitrary coarse grainings because they are not imposed by the external
observer but are intrinsic properties of the system. To our knowledge 
only three types of such intrinsic coarse grainings have been proposed. 
Namely, projections onto the $K$-partition~\cite {Prig7,Misra}, onto
the generating partition~\cite {Gustaf} of Kolmogorov dynamical systems
and onto Markov partitions~\cite {Nicolis}.

We shall not discuss further this interesting subject related also to the
symbolic representation of dynamics~\cite {ref15} but consider
arbitrary coarse grainings as so far no a priori reason to decide over
the proper natural partition has been proposed to our knowledge.

The influence of the coarse graining to the approach to equilibrium will 
be studied through the decay rates of the autocorrelation functions of 
the observable phase functions $f$:
\begin{equation}
C^{(n)}(f)=\frac{\int_{X}(V^{n}f)(x)f(x)d\mu (x)}{\int_{X}f^{2}(x)d\mu (x)}.
\label {corfun}
\end{equation}
The coarse grained autocorrelation function $C_{P}^{(n)}(f)$ is 
\begin{equation}
C_{P}^{(n)}(f)=\frac{\int_{X}(V_{P}^{n}f)(x)Pf(x)d\mu (x)}
{\int_{X}(Pf(x))^{2}d\mu (x)}.  
\label {corfunP}
\end{equation}
However, in the following we shall see that the autocorrelation functions
are not sensitive enough to discriminate results of different approaches. 
In order to analyze a system in more detail, we introduce the decay rate
$\tau ^{(n)}$ at time (stage) $n$:
\begin{equation}
\tau^{(n)}(f)=-\log \frac{C^{(n+1)}(f)-C^{(n)}(f)}{C^{(n)}(f)-C^{(n-1)}(f)}.
\end{equation}
This definition is motivated by the following observation. If the
autocorrelation function decreases (or increases) monotonically at the
points $n-1$, $n$, $n+1$, it can be written as
\begin{equation}
C^{(n)}=A(n)+B(n)\exp (-\kappa (n)n)  \label{repr6}
\end{equation}
at these points. In this case the decay rate is $\tau ^{(n)}=\kappa (n)$.
As we can usually expect that the representation (\ref{repr6}) is the
leading term of the asymptotics for sufficiently large $n$ and that 
$A(n)$, $B(n)$, and $\kappa (n)$ become independent of $n$ for large $n$, 
the decay rate $\tau^{(n)}$ converges to the decay rate of the system 
$\kappa (\infty)$. Physically speaking, in most cases one needs information
about the widths and the lifetimes of the system, i.e. $\tau^{(n)}$, rather
than just the decaying profile $C^{(n)}$.

\section {Coarse graining the Baker transformation}

We shall apply the coarse graining to the Baker transformation
defined~\cite {Bakerdef} on the torus $[0,1]\times [0,1]$ by the formula
\begin{equation}
B(x,y)=\left\{
\begin{array}{ll}
(2x,\frac{y}{2}), & {\rm for\ \ }0\leq x\leq 1/2 \\
(2x-1,\frac{y+1}{2}), & {\rm for\ \ }1/2\leq x\leq 1
\end{array}
.\right.
\end{equation}
In order to study the applicability of the coarse graining for this
transformation, we introduce two different partitions $\zeta ^{s}$ 
and $\zeta ^{t}$ such that 
$\mu(Z_{i}^{s})=\mu (Z_{j}^{t})$ for all $i$, $j$. As these
partitions have cells of the same measure, the role of geometry is
manifested in the clearest way. It is worth noticing here that we cannot 
use for this study an one-dimensional map as we cannot choose two different
partitions with the same measure in one dimension. In Fig.~1, we present
both partitions for the number of the cells in each direction $M=4$.

The calculation of autocorrelation functions (\ref{corfun},\ref{corfunP})
involves integration over each cell. The numerical realization of this
integration results in loss of accuracy. To avoid this problem, we use 
the fact that the function $f^{(n)}=(PBP)^{n}f$ is piece-wise constant for
$n\geq 1$. For such functions, the successive iterations can be written as
\begin{equation}
f_{k}^{(n+1)}=\frac{1}{\mu (Z_{k})}\sum_{i}f_{i}^{(n)}\mu (Z_{k}\bigcap
B(Z_{i})).  
\label {iterat}
\end{equation}
It is worth noticing that the latter representation is rather effective 
from the computational point of view as the sum in Eq.~(\ref{iterat})
involves very few terms. The only remaining integration in
Eq.~(\ref{corfunP}) is the calculation of $Pf$. We use the rectangular
quadrature formula for this integration. The numerical investigation 
shows that the number of integration points 40 by 40 for each
cell is enough to reach the convergence.

Now, after the description of the method, we present the results. We start
with the initial function
\[
f(x,y)=x+y.
\]
With this particular choice, it is possible to calculate the autocorrelation
function explicitly. A straightforward analytical calculation gives:
\begin{equation}
C^{(n)}=\frac{6}{7}(1+\frac{1}{8}n 2^{-n}+\frac{1}{6} 2^{-n}).
\label {Bakercorr}
\end{equation}
It is interesting to point out that the second term in 
expression~(\ref{Bakercorr}) corresponds to the Jordan block of the 
second order in the generalized spectral decomposition~\cite {Bakerspectr}.
With expression (\ref{Bakercorr}), the decay rate can be obtained exactly:
\begin{equation}
\tau ^{(n)}=\log 2-\log \left( 1+\frac{3}{3n-2}\right) {.}
\label {Bakerrates}
\end{equation}
The exact expressions (\ref{Bakercorr},\ref{Bakerrates}) are used for the
comparison with the coarse graining results.

The autocorrelation function and the local decay rates for both
partitions with $M=200$, and analytical results are presented in Fig.~2. 
The first conclusion we can make here is that coarse graining reproduces 
the correlation function with accuracy better than $10^{-4}$. Hence, if 
one is interested only in similar integral characteristics, coarse 
graining can be used.

However, in many cases one needs more detailed characteristics of the
evolution, for example the decay rates. Here the situation is changed as
different partitions result in significantly different  decay rates. 
While in this particular case the square partition $\zeta^{s}$ produces 
results that are surprisingly close to the exact ones, the results for 
the triangle partition $\zeta ^{t}$ differ drastically although they agree 
up to 11 iterations.

The decay rates for the partitions with $M=800$ are presented in Fig.~3.
For these partitions the results agree very well in a wider
region, up to about 20 iterations. However, further iterations again show
huge disagreement between the results for the triangle partition and the
exact results.

The square partitions may also give irrelevant results. Namely, if we
consider the rectangle partitions with $2^N$ subdivisions of the $x$ 
axis and $2^M$ subdivisions of the $y$ axis, than after $N+M-1$ iterations
the coarse grained correlation function reaches equilibrium exactly. 
Hence, there is no decay after this number of stages. This statement is 
also true for the square partition when $N=M$. However, there exists
rectangle partition with the same cell area that results in the proper 
decay curve. We illustrate this discussion in Fig.~4. Therefore the decay
rates are very sensitive to the choice of the type of the partition.

One may expect that despite the big variations of the decay
rates, an average of them might be stable. To analyze this
possibility, we present in Fig.~5 the average decay rates
$$
\tau _{{\rm av}}^{(n)}=\frac{1}{10}\sum_{i=n+1}^{n+10}\tau ^{(i)}
$$
for $n=10$ and $n=20$. However also in this case the previous conclusions
remain valid: while the square partition produces decay rates robust with
respect to changes of the cell area, the triangle partition gives results 
which change irregularly with respect to changes of the cell area. 
However, the average decay rate $\tau_{{\rm av}}^{(10)}$
for the triangle partition is stabilized when $M$ is rather big. But we
cannot restrict ourselves to this small number of iterations ($n \sim 10$) 
as the decay rates reach their asymptotics much later. The results 
for the $\zeta^{s}$ show this clearly as the difference of 
$\tau_{{\rm av}}^{(10)}$ and $\tau_{{\rm av}}^{(20)}$ is rather pronounced.

We have already mentioned that our results do not depend on the choice of 
the observable function $f(x,y)$. To illustrate this, we present in Fig.~6
the autocorrelation function and the local decay rates for both partitions
with $M=800$, and the initial function $f(x,y)=x\sqrt{x+y}$. Analytical
results are not available for this function. One can see that the results
have the same qualitative behavior as previous ones. Calculations for 
other initial functions also give the similar results, so our conclusions 
are valid and independent of the initial function.

\section {Coarse graining the periodic torus automorphisms}

In the previous section we discussed coarse graining applied for the Baker
transformation. One could see that, despite the problems with the decay
rates, the method reproduces the autocorrelation functions rather well. 
Here we shall show that this is not always the case, and coarse graining
applied to some dynamical systems may produce even qualitatively wrong
autocorrelation functions.

Let us consider the periodic automorphisms $T$ of the torus
$[0,1] \times [0,1]$:
\[
T\left(
\begin{array}{l}
x \\
y
\end{array}
\right) =\left(
\begin{array}{l}
(T_{11}x+T_{12}y)~{\rm mod}\ 1 \\
(T_{21}x+T_{22}y)~{\rm mod}\ 1
\end{array}
\right) .
\]
The coefficients $T_{ij}$ are integers with the property
\begin{equation}
T^{k}=I \qquad {\rm for\ some\ } k \in N.
\end{equation}

There exist few families of periodic torus automorphisms $T$ with 
only possible periods $k=2$, 3, 4, and 6 (Appendix 1). As described in 
the previous section, we construct the autocorrelation functions
(\ref{corfun},\ref{corfunP}). Due to the periodicity of the
transformation $T$, the exact autocorrelation function is also 
periodic. However, the coarse grained autocorrelation functions may 
show damping as a result of information loss. We present two such 
examples in Fig.~7 . For the transformation
$$
T=\left(
\begin{array}{rr}
1 & 2 \\ -1 & -1
\end{array}
\right) 
$$ 
we used the partition $\zeta^{s}$ with $M=20$. This type of the partition
gives reasonable results for the Baker transformation. Here, however, 
one can see that the autocorrelation function approaches equilibrium
as if there was damping. Hence, the periodicity is not preserved. 
Moreover, for the simple transformation 
$$
T=\left(
\begin{array}{rr}
0 & -1 \\ 1 & 0
\end{array}
\right) 
$$ 
there exist also partitions giving rise to damping. As an example, 
we plot in Fig.~7 the autocorrelation function for the rectangle 
partition with 5 subdivisions for the $x$ axis and 7 subdivisions for
the $y$ axis.

\section {Conclusions}

For the Baker transformation, coarse graining reproduces the shape of
the autocorrelation functions (Figs.~2,6), but decay rates are hardly
reproduced (Figs.~2-6). Moreover, some partitions like in Fig.~4 give 
rise to exact approach of equilibrium after a finite time while 
others may give the correct asymptotic behaviour. 
For the periodic torus automorphisms, coarse graining introduces 
artificial damping of the autocorrelation functions.

The above mentioned conclusions illustrate once more the subjectivistic
character of coarse graining. Therefore when employed, coarse graining 
should be chosen with care, for example by selecting partitions 
intrinsic to the dynamics, as discussed at the end of section~2. 
In fact the answer to the inverse problem 
of statistical physics is~\cite {Gustaf} that all stationary Markov 
processes arise as exact projections onto the generating partitions of 
Kolmogorov dynamical systems, in the spirit of Misra-Prigigine-Courbage
theory of irreversibility~\cite {Prig7,Misra}. In fact such intrinsic 
partitions are not defined by any observations but they are objective
properties of the dynamical evolution.

Concerning the general issue of irreversibility, the possibilities opened 
by extending the dynamical evolution or by intertwining the dynamical
evolution with Markov processes~\cite {Prig7,Prig8,Prig9} are a challenging
physical and mathematical research direction.

\vskip5mm
{\bf Acknowledgements}

Helpful discussions with I.~Prigogine, S.~Shkarin and Z.~Suchanecki
are gratefully acknowledged. The financial support of the Belgian 
Government through the Interuniversity Attraction Poles and the 
National Lottery of Belgium is gratefully acknowledged.

\newpage
\begin{center}
\bf \large Appendix 1. The periodic torus automorphisms.
\end{center}

The periodic torus automorphisms 
$$
T=\left(\begin{array}{rr} a & b \\ c & d \end{array} \right)
$$ 
with integer coefficients satisfy the equation
$$ T^k = I, $$
where $k=2,3,\ldots $ is the period. As $|\det{T}| = 1$, we have 
$\det{T} = \pm 1$. Let us transform the matrix $T$ to the diagonal or 
the Jordan form with the unitary transformation. As the identity matrix 
is unchanged, we can conclude that the eigenvalues of the matrix $T$ 
satisfy $|\lambda_{1,2}|=1$. Let us examine all possible cases 
corresponding to real and complex eigenvalues.

I. The eigenvalues are real. In this case we have three possibilities: \\
1. $\lambda_1=\lambda_2=1$. As the Jordan block cannot be unitarely
equivalent to the identity matrix, the only possible transformation 
is the trivial one, $T=I$. \\
2. $\lambda_1=\lambda_2=-1$. As in the first case, the only possible 
transformation is $T=-I$ which has the period $k=2$. \\
3. $\lambda_1=1$, $ \lambda_2=-1$. In this case we have a family of 
periodic automorphisms. In order do describe them, we may use two 
invariants: the determinant $\det T = -1$ and trace ${\rm tr} T = 0$, 
and write down the family as 
\beq
\label {famfirst}
\left( \begin{array}{rr} a & b \\ c & -a \end{array} \right)
\quad \mbox{with} \quad \det T = -a^{2}-bc=-1.
\eeq
These automorphisms have the period $k=2$.

II. The eigenvalues are complex. As the matrix has real coefficients, 
the eigenvalues are conjugated to each other and can be written as 
$\lambda_1=e^{ i \varphi}$, $\lambda_2=e^{- i\varphi}$, $\varphi \in R$. 
Using the invariants of the matrix, $\det T=1$ and 
${\rm tr} T = 2\cos\varphi$, we may write the 
corresponding family as 
$$
\left( \begin{array}{rr} a & b \\ c & 2\cos\varphi-a \end{array} \right)
\quad \mbox{with} \quad \det T = a(2\cos\varphi-a)-bc=1. 
$$
The cases when $\varphi=0$ and $\varphi=\pi$ give in fact real 
coefficients and are already analyzed. The 
only remaining cases when the matrix has integer coefficients are: \\
1. $\cos\varphi=0$, $\varphi=\pi/2$. Here we have the family 
\beq
\left( \begin{array}{rr} a & b \\ c & -a \end{array} \right)
\quad \mbox{with} \quad \det T = -a^{2}-bc=1. 
\eeq
These automorphisms have the period $k=4$. \\
2. $\cos\varphi=1/2$, $\varphi=\pi/3$. The corresponding family is
\beq
\left( \begin{array}{rr} a & b \\ c & 1-a \end{array} \right)
\quad \mbox{with} \quad \det T = a(1-a)-bc=1. 
\eeq
These automorphisms have the period $k=6$. \\
3. $\cos\varphi=-1/2$, $\varphi=2\pi/3$. The corresponding family is
\beq
\label {famlast}
\left( \begin{array}{rr} a & b \\ c & -1-a \end{array} \right)
\quad \mbox{with} \quad \det T = -a(1+a)-bc=1. 
\eeq
These automorphisms have the period $k=3$.

All possible periodic torus automorphisms are given by 
(\ref{famfirst}) -- (\ref{famlast}).

\newpage
\begin{center}
Figure captions
\end{center}

Fig. 1. The partitions used in calculations for $M=4$. Fig. 1a shows 
the square partition $\zeta ^{s}$, and Fig. 1b shows the triangle partition $\zeta^{t}$.

Fig. 2. The correlation function $C^{(n)}$ (the solid line) and decay rates
$\tau ^{(n)}$ for squares (the long-dashed line), triangles (the dot-dashed
line), and analytical results (the short-dashed line). The number of
cells is 200 by 200.

Fig. 3. Decay rates $\tau ^{(n)}$ for squares (the short-dashed line),
triangles (the long-dashed line), and analytical results (the solid line).
The number of cells is 800 by 800.

Fig. 4. Decay rates $\tau ^{(n)}$ for squares (the dashed line) with 
number of cells 256 by 256 (the cell area $1.52588\ 10^{-5}$), for 254 by 
258 rectangles (the dot-dashed line, the cell area $1.52597\ 10^{-5}$), 
and analytical results (the solid line). The second and the third 
lines practically coincide.

Fig. 5. Average values $\tau _{{\rm av}}^{(n)}$ for $n=10$ and $n=20$ as 
the function of number of cells M for squares and triangles. The solid 
line is for $\tau _{{\rm av}}^{(10)}$ for triangles, the long-dashed line 
is for $\tau _{{\rm av}}^{(20)}$ for triangles, the short-dashed line is 
for $\tau _{{\rm av}}^{(10)}$ for squares, and the dot-dashed line is 
for $\tau _{{\rm av}}^{(20)}$ for squares.

Fig. 6. The correlation function $C^{(n)}$ (the solid line) and decay 
rates $\tau ^{(n)}$ for squares (the long-dashed line) and triangles 
(the short-dashed line). The initial function is $f(x,y)=x\sqrt{x+y}$. 
The number of elements is 800 by 800.

Fig. 7. The correlation function $C^{(n)}$ for the periodic torus
automorphisms $T=\left(
\begin{array}{rr}
1 & 2 \\
-1 & -1
\end{array}
\right)$ with the partition $\zeta ^{s}$ and $M=20$ (line 1), and for the
$T=\left(
\begin{array}{rr}
0 & -1 \\
1 & 0
\end{array}
\right) $ with the partition of 5 intervals for $x$ and 7 intervals 
for $y$ (line 2).

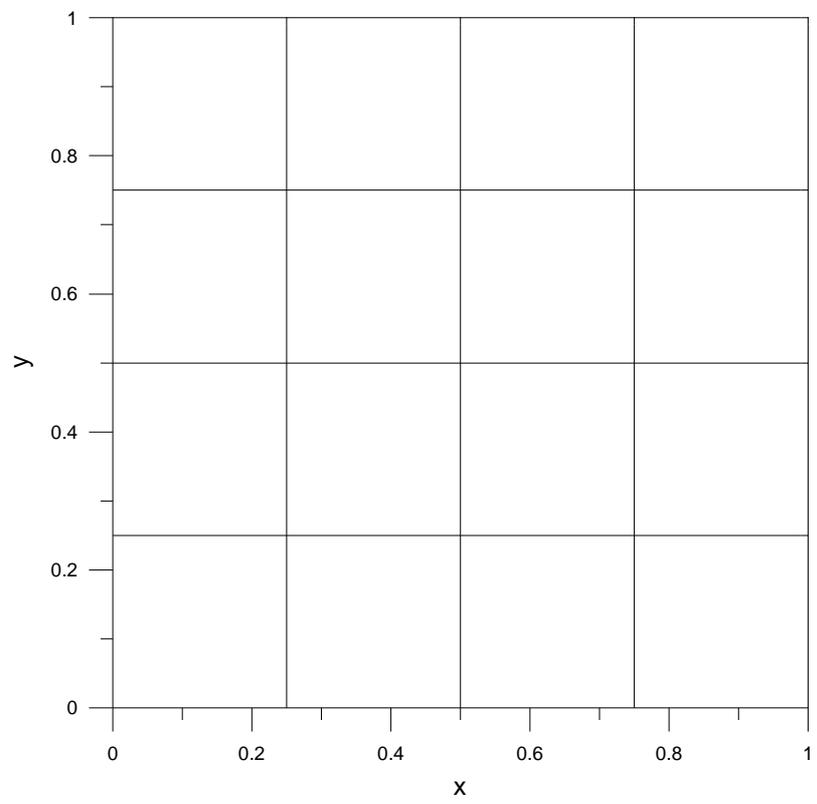
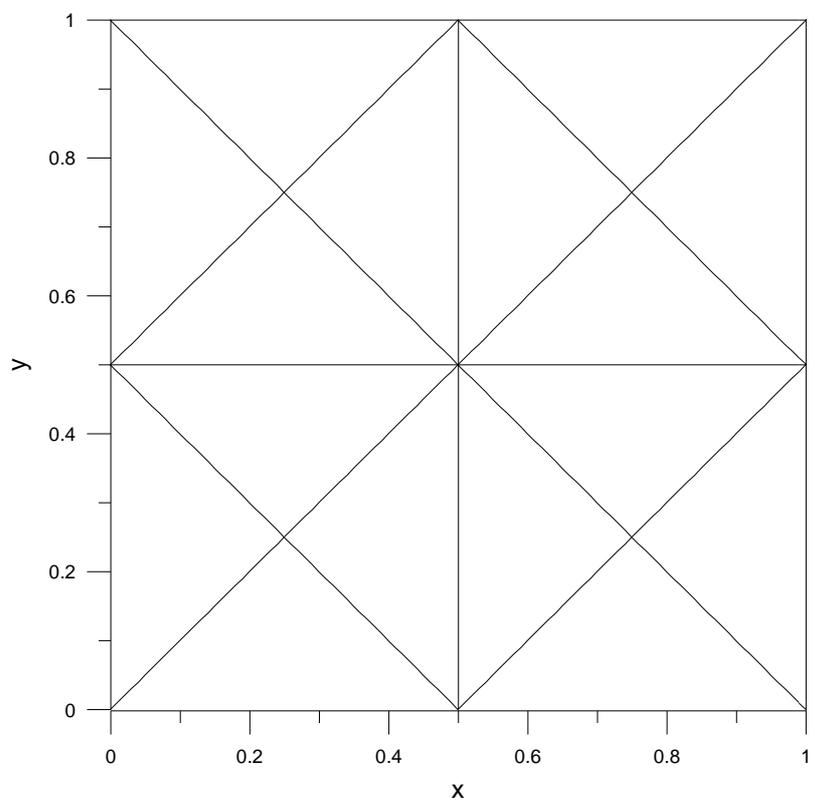
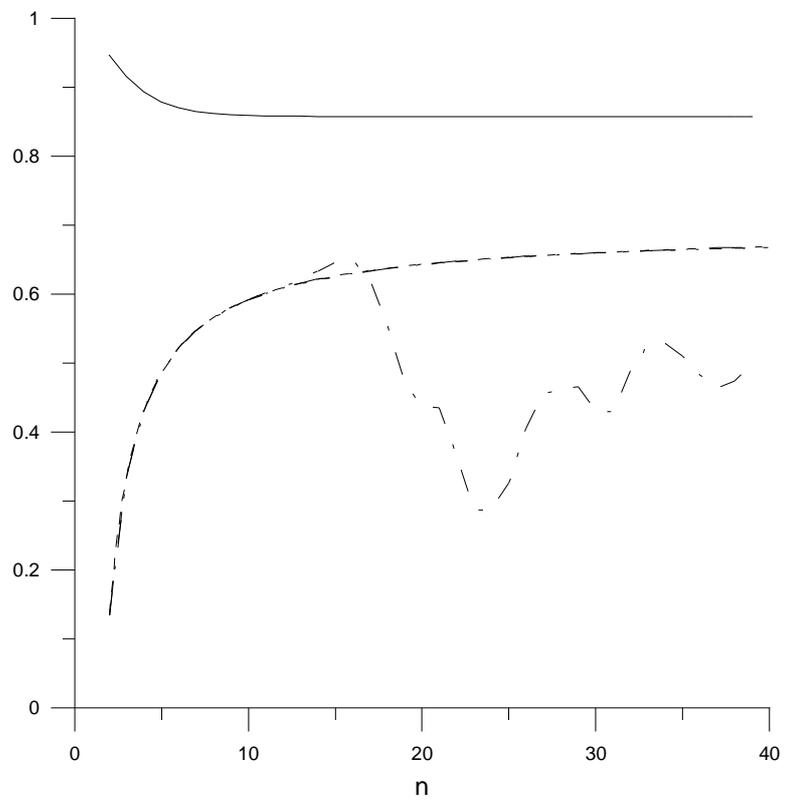
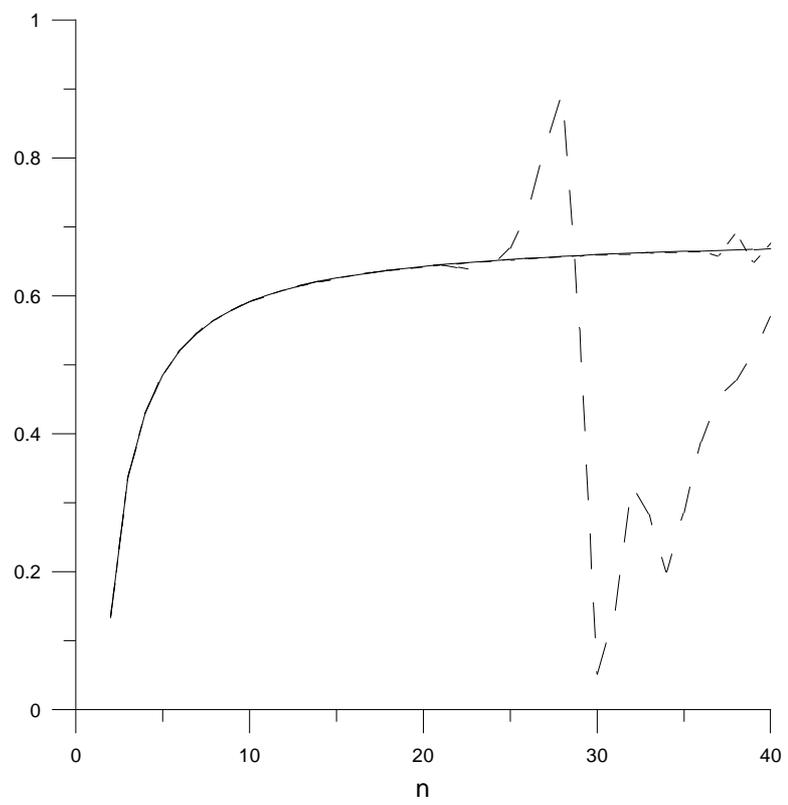
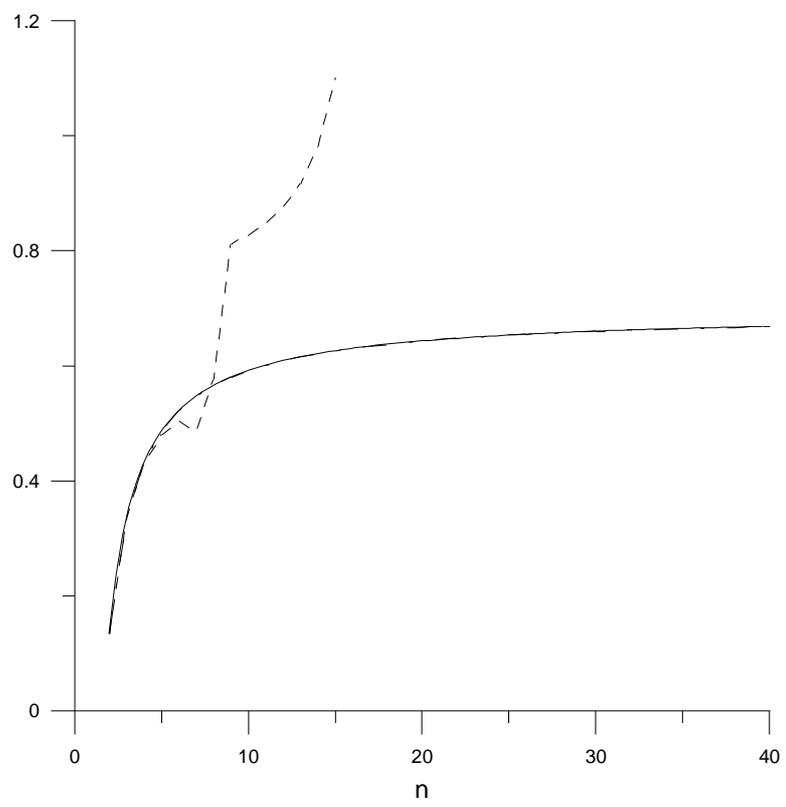
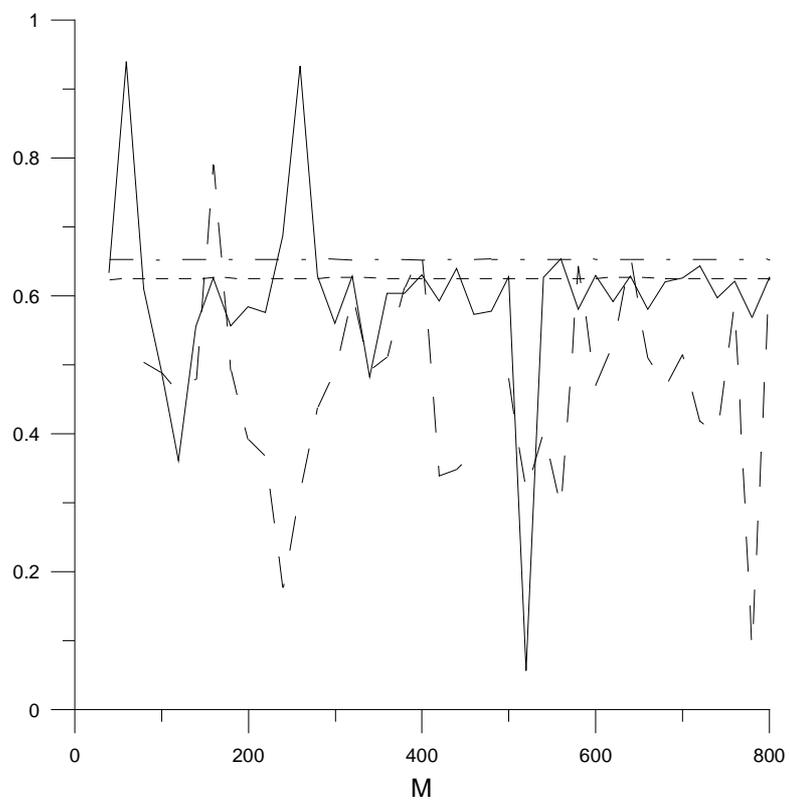
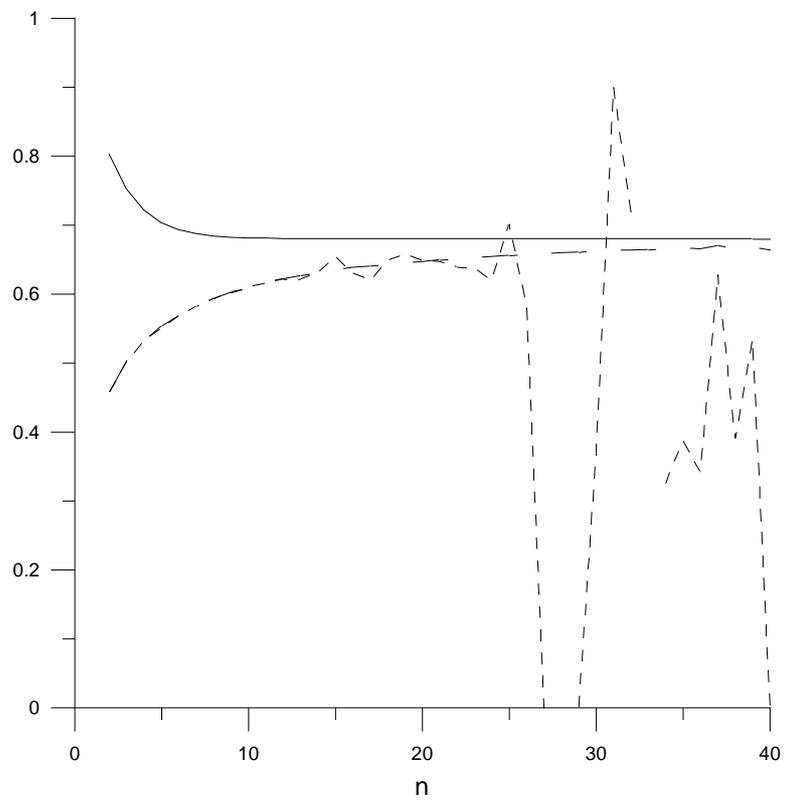
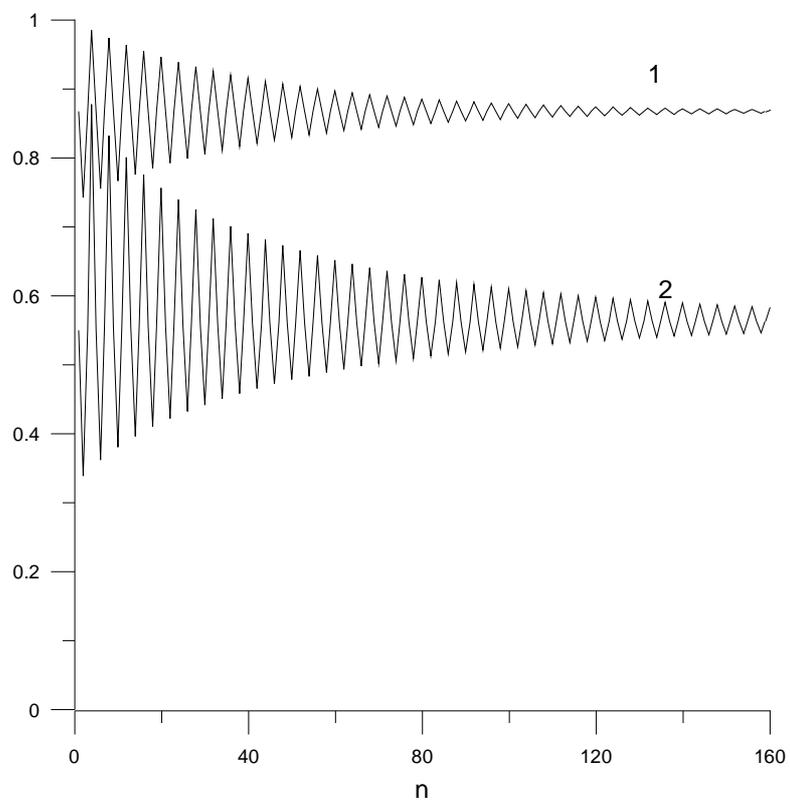
\end{document}